%% file: main.tex
\documentclass[fleqn,usenatbib]{mnras}

\pdfoutput=1
\usepackage{graphicx}
\usepackage{float}
\usepackage{amssymb,amsmath}
\usepackage[T1]{fontenc}
\usepackage{bm}
\usepackage{xcolor}
\usepackage[normalem]{ulem}
\usepackage{array}
\usepackage{newtxtext}

\newcolumntype{L}[1]{>{\raggedright\let\newline\\\arraybackslash\hspace{0pt}}m{#1}}
\newcolumntype{C}[1]{>{\centering\let\newline\\\arraybackslash\hspace{0pt}}m{#1}}
\newcolumntype{R}[1]{>{\raggedleft\let\newline\\\arraybackslash\hspace{0pt}}m{#1}}

\newcommand{\lr}[1]{\left({#1}\right)}
\newcommand{\dd}{\ensuremath{\mathrm{d}}}%
\newcommand{\unit}[2]{#1\,\mathrm{#2}}

\newcommand{\ut}[3]{#1^{+#2}_{-#3}}
\newcommand{\ex}[1]{\exp\left({#1}\right)}
\newcommand\thefontsize[1]{{#1 The current font size is: \f@size pt\par}}
\newcommand{\R}{{\mathcal{R}}}%

\title[Constraints on low diffusion zones]{Unstable cosmic-ray nuclei constrain low-diffusion zones in the Galactic disk}
\author[H. Jacobs et al.]{Hanno Jacobs,$^{1}$\thanks{E-mail: jacobs@physik.rwth-aachen.de} Philipp Mertsch,$^{1}$\thanks{E-mail: pmertsch@physik.rwth-aachen.de} and Vo Hong Minh Phan$^{1}$\thanks{E-mail: vhmphan@physik.rwth-aachen.de}\\
$^{1}$ Institute for Theoretical Particle Physics and Cosmology (TTK), RWTH Aachen University, 52056 Aachen, Germany
}
\date{\today}

\begin{document}
\label{firstpage}

\maketitle

\begin{abstract}
Observations of the vicinity of a variety of galactic gamma-ray sources have indicated a \emph{local} suppression of diffusivity of cosmic rays by up to three orders of magnitude. However, the impact of these low-diffusion zones on \emph{global} properties of cosmic-ray transport is however only poorly understood. Here, we argue that cosmic-ray nuclear ratios, like the boron-to-carbon ratio and relative abundances of Beryllium isotopes are sensitive to the filling fraction of such low-diffusion zones and hence their measurements can be used to constrain the typical sizes and ages of such regions. We have performed a careful parameter study of a cosmic-ray transport model that allows for different diffusion coefficients $\kappa_{\mathrm{disk}}$ and $\kappa_{\mathrm{halo}}$ in the galactic disk and halo, respectively. Making use of preliminary data from the AMS-02 experiment on the ratio of Beryllium isotopes, we find a $3.5 \sigma$ preference for a suppression of the diffusion coefficient in the disk with a best-fit value of $\kappa_{\mathrm{disk}}/\kappa_{\mathrm{halo}} = \ut{0.20}{0.10}{0.06}$. We forecast that with upcoming data from the HELIX balloon experiment, the significance could increase to $6.8 \sigma$. Adopting a coarse-graining approach, we find that such a strong suppression could be realised if the filling fraction of low-diffusion zones in the disk was $\sim 66 \, \%$. We conclude that the impact of regions of suppressed diffusion might be larger than usually assumed and ought to be taken into account in models of Galactic cosmic ray transport.
\end{abstract}


\input{sections/Introduction}
\input{sections/Model}
\input{sections/Results}
\input{sections/Discussion}

\input{sections/SummaryConclusion}

\section*{Acknowledgements}

The authors would like to thank Jakob Böttcher for advice on likelihood ratio tests in the presence of parameter bounds and acknowledge helpful conversations with Dieter Breitschwerdt, Stefano Gabici, Yoann Génolini and Pierre Salati.

\bibliographystyle{mnras}
\bibliography{cheese}
\newpage
\appendix
\newpage
\onecolumn
\input{sections/Appendix}

\end{document}

%% file: sections/Introduction.tex
\section{Introduction}
One of the cornerstones of the standard picture of cosmic ray (CR) transport (see e.g. \citealt{strong2007,grenier2015,Gabici2019} for a review) is related to the overabundances of certain, so called \emph{secondary} species with respect to solar system values. 
Such a difference is commonly interpreted as resulting from spallation reactions of \emph{primary} CRs with interstellar gas. 
The measurement of secondary-to-primary ratios like B/C therefore allows us to constrain the matter traversed during propagation, called the grammage. 
As the amount of grammage accumulated is about three orders of magnitude larger than the column density of the disk, CRs must propagate diffusively. 
In fact, flux ratios involving unstable secondaries such as $\mathrm{^{10}Be}$ allow us to infer that CRs are confined in a larger region called the Galactic halo of CRs, or halo for short, extending to kiloparsec distances above and below the disk. 
All these interpretations can be formalised in a standard framework where CRs are produced in the Galactic disk and diffuse in the halo by scattering upon turbulent Galactic magnetic field, first proposed by \citet{Ginzburg1964}. 
The combination of B/C and $\mathrm{^{10}Be}$ then allows us to determine the two most important parameters of this model which are the diffusion coefficient $\kappa$ and the halo height $H$~\citep{2002A&A...381..539D,2022arXiv220413085L}.

As magnetic turbulence controls both $\kappa$ and $H$, it is essential for a comprehensive picture of Galactic CR propagation. 
Yet, the question of how magnetised turbulence is generated and distributed on large scales remains open. 
Turbulence can be generated both by supernova explosions and by inhomogeneities of CR density via the CR \textit{streaming instability}~\citep{kulsrud1969,skilling1975c,marcowith2021}. 
For Galactic propagation, the former is believed to be particularly important for CR particles with high rigidity above a few hundred GV while the latter might be more relevant for CRs below a few hundred GV~\citep{blasi2012}. This has been employed to explain the change in the rigidity dependence of $\kappa$ at $\R \simeq 300$ GV as inferred from B/C data~\citep{genolini2017}. 
We note, however, that most of the analyses adopt the simplified assumption that $\kappa$ is homogeneous, specifically that it is the same in the disk and in the halo (see, however \citealt{2012ApJ...752L..13T,2014ApJ...782...36E}).
More recent studies suggest that the spatial dependence of $\kappa$ might be more complicated if turbulence self-generated via the CR streaming instability is taken into account \citep{recchia2016,Evoli2018}.     

In fact, observations seem to indicate that the diffusion coefficient could be different between the halo and the disk, at least for certain regions in the disk. 
Gamma-ray observations of potential CR sources such as supernova remnants or pulsars have been interpreted to have regions with a suppressed isotropic diffusion coefficient in their surroundings. 
For example, HAWC~\citep{Abeysekara2017}, \textit{Fermi}-LAT~\citep{DiMauro2019} and H.E.S.S.~\citep{Mitchell2022} observed halos of gamma rays (referred to as gamma-ray halos to be distinguished from the Galactic halo of CRs) at GeV and TeV energies around the pulsars Monogem and Geminga. 
These halos originate from high-energy e$^+$/e$^-$ pairs accelerated by the pulsars up to energies of $\unit{100}{TeV}$~\citep{Abeysekara2017} and produce gamma rays via inverse Compton scattering upon the cosmic microwave background and interstellar radiation fields. 
These observations provide some support for the hypothesis that pulsars are sources of primary positrons, capable of explaining the positron excess observed by AMS-02~\citep{Aguilar2013} and PAMELA~\citep{Adriani2010}. 
Interestingly, the radial profile of gamma-ray halos cannot be explained by models where electrons diffuse isotropically from pulsars with a diffusion coefficient as inferred by fits to local CR data, but they require a diffusion coefficient lower by about two to three orders of magnitude~\citep{Abeysekara2017}. 
Under the initial assumption that the diffusion coefficient is that low in the entire disk, electrons at TeV energies could not reach Earth, since the diffusion time would be larger than the energy loss time. 
However, many experiments observe these electrons \citep[e.g.][]{DAMPE2017,CALET2018}, hence suppression of the diffusion coefficient by that order can be excluded~\citep{Hooper2018}. 
In order to simultaneously explain the positron excess and the gamma-ray halos a two-zone diffusion model was introduced~\citep{Tang2019} with inhibited diffusion in a spherical region at least the size of the gamma-ray halo and the usual interstellar medium value outside. 
Within this two-zone approach, it is possible to explain the positron excess with a distribution of pulsars~\citep{Manconi2020}. 
However, the origin of these low-diffusion regions remains uncertain. 
(See, however, \citealt{Evoli2018Halo}). 

Additionally, observations of gamma-rays from dense molecular clouds near supernova remnants can be interpreted with a suppressed diffusion coefficient in this area~\citep{Gabici2010}. 
In this scenario, hadronic CRs accelerated by the supernova remnant escape into the interstellar medium and propagate to nearby dense molecular clouds, where they produce gamma-rays via pion production. 
The diffusion coefficient required to explain the intensity of gamma-rays is suppressed by around two orders of magnitude with respect to the Galactic average. 
This suppression can be self-generated via CR streaming instability~\citep{Malkov2013,Nava_2016,Nava_2019,Brahimi2020,Recchia_2021} and could last for several $\unit{100}{kyr}$ at GeV energies~\citep{Jacobs2022}. 
Furthermore, suppression of the diffusion coefficient at GeV energies has been inferred in dense molecular clouds and in stellar bubbles~\citep{Aharonian2019,Yang2023}.
Hence, there is plenty of evidence of inhomogeneous diffusion and the impact of these low diffusion zones on Galactic cosmic ray propagation might be non-negligible. 

Here, we will devise a simple semi-analytical model of CR propagation in a scenario where we have a suppressed averaged diffusion in the Galactic disk and investigate constraints of upcoming AMS-02~\citep{Wei2022} and HELIX~\citep{Park2019} $\mathrm{^{10}Be/^9Be}$ measurements. 
We then show how to calculate a large-scale averaged diffusion coefficient from smaller-scale features using the numerical solution of a stochastic differential equation. Finally, we explain the implications of our findings and conclude.
Detailed calculations and additional results are presented in a number of appendices. 

%% file: sections/Model.tex
\section{Model}

\begin{figure*}
    \centering
    \includegraphics[width=\textwidth]{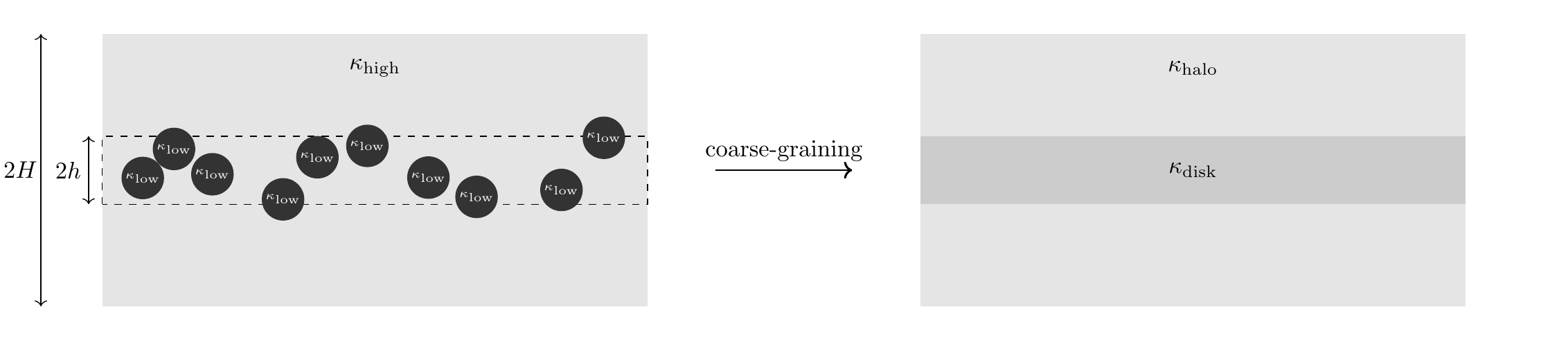}
    \caption{ 
    Schematic sketch of low diffusion zones in the disk and the effect of course-graining. Zones of low diffusivity $\kappa_{\mathrm{low}}$ in the Galactic disk of height $2h$ are surrounded by a zone with high diffusivity $\kappa_{\mathrm{high}}$, which extends into the Galactic halo of height $2H$. Coarse-graining leads to an averaged diffusion zone in the disk $\kappa_{\mathrm{disk}}$ while in the halo $\kappa_{\mathrm{halo}}=\kappa_{\mathrm{high}}$.}
    \label{fig:coarse-graining}
\end{figure*}

We use a $1D$ transport equation to describe CR propagation in the Galaxy following~\citet{Parker1965,Ginzburg1964}. 
Cosmic rays are assumed to propagate within a magnetised halo of size $2H$ extending symmetrically above and below the Galactic disk. The halo is assumed to have a negligible gas density and a free escape boundary condition is imposed at the halo's boundaries. 
The sources of CRs are considered to be located only within the disk of height $h\ll H$. Similarly, the interstellar gas, which is the target of interactions inducing energy loss for CRs, is confined to this region.
Since all observed and potential sources of suppressed diffusion are confined to the disk, we distinguish between two zones, the disk and the halo in the following.

Analytically solving the transport equation for a diffusivity with randomly distributed regions of suppression is intractable. 
A numerical solution with the finite-difference technique that is commonly used in CR transport codes is also unfeasible: Due to the large dynamical range in diffusivity, unrealistically small times steps would be required for the algorithm to remain accurate. 
Here, we therefore resort to a coarse-graining approach where we adopt a spatially homogeneous diffusion coefficient $\kappa_{\mathrm{disk}}(\R)$ in the disk which is related to an unsuppressed diffusion coefficient $\kappa_{\mathrm{high}}(\R)$ by a scaling factor $\alpha$,
\begin{equation}
\kappa_{\mathrm{disk}}(\R) = \alpha\kappa_{\mathrm{high}}(\R) \quad \text{with} \quad 0\leq\alpha\leq 1 \, .
\end{equation}
Here, we have assumed the same rigidity-dependence for $\kappa_{\mathrm{disk}}(\R)$ and $\kappa_{\mathrm{high}}(\R)$. 
This can be motivated, e.g. for gamma-ray halos around PWNe, by the fact that the rigidity-dependence in the low-diffusion zone has been found to be compatible with a Kolmogorov phenomenology that is commonly assumed for $\kappa_{\mathrm{high}}$~\citep{DiMauro2019}. 
We note that this approach is also flexible enough to cover other potential sources of suppressed diffusion, like CR self-confinement, as long as the suppressed and unsuppressed diffusion coefficients have the same rigidity-dependence. 
We posit that this scenario deserves investigation on its own merits. 
For the scenario where $\alpha$ results from the presence of bubbles of suppressed diffusion, we provide some relations between the ratio $(\kappa_{\text{low}} / \kappa_{\text{high}})$, the filling fraction $f$ of bubbles in the disk and the resulting $\alpha$ in Sec.~\ref{sec:discussion}.

In the halo, the absence of low diffusion zones implies \mbox{$\kappa_{\mathrm{halo}}=\kappa_{\mathrm{high}}$}.
The rigidity dependence of the diffusion coefficient is modelled as a broken power law to account for spectral breaks observed in CR data at low~\citep{Vittino2019} and high rigidities~\citep{Evoli2019_AfterAMS}, 
\begin{eqnarray}
    \kappa(z,\R)&=&\beta\left( 1+\lr{\frac{\R}{\R_l}}^{-\frac{1}{s_l}} \right)^{s_l(\delta-\delta_l)} \left( \frac{\R}{10\,{\rm GV}}\right)^{\delta}\nonumber\\
    &\times&\left(1+\lr{\frac{\R}{\R_h}}^{\frac{1}{s_h}}\right)^{s_h(\delta_h-\delta)} \times
\begin{cases}
\kappa_{\mathrm{disk}} &\text{$z\leq h$\, ,}\\
\kappa_{\mathrm{halo}} &\text{$z>h$\, .}
\end{cases}
\end{eqnarray}
Some of the parameters have little effect on our results and so we consider values which have been found in other studies. Specifically, we fix $s_l=0.04$, which has minimal impact on the results, as shown by~\cite{Weinrich2020}. 
Furthermore, the high energy break parameters can be determined by primary nuclei~\citep{Evoli2019_AfterAMS}: $s_h=0.5, \ \delta_h=0.34, \ \R_h=\unit{312}{\mathrm{GV}}$. 
All other parameters remain as free parameters.

In the steady state, the transport equation for Galactic CRs reads~\citep{Berezinskii1990}:
\begin{eqnarray}
        &&\frac{\partial }{\partial z}\lr{v\psi_j-\kappa(z,\R)\frac{\partial \psi_j}{\partial z}} +\frac{\partial}{\partial p}\left[\lr{\frac{\dd p}{\dd t}}\psi_j-\frac{p}{3}\frac{dv}{dz} \psi_j\right] \nonumber \\
        &+&\frac{1}{\gamma \tau_j}\psi_j + 2h\delta(z)\beta c n_{\mathrm{gas}}\sigma_j \psi_j = Q_{j}(p) \, , \label{eqn:TPE}
\end{eqnarray}
where $\psi_j$ is the density differential in momentum of CRs of species $j$, connected to the phase space density $f_j$ by $\psi_j=4\pi p^2 f_j$. 
The terms on the right hand side describe advection, diffusion, (adiabatic) energy losses, decay with boosted decay time $\gamma\tau_j$ and spallation of CRs, respectively. 
While we distinguish between diffusion at $z<h$ and $z>h$, we neglect the spatial distribution of spallation, energy losses and the sources in the disk, since we expect the effect to be small and approximate them with a delta distribution $\delta(z)$. 
We make use of the cross sections provided by ~\citet{Evoli2019_AfterAMS,Evoli2020_Be}.
The advection speed $v(z)=\mathrm{sgn}(z)v_c$ is directed vertically out of the Galactic disk and left as a free parameter, giving $\dd v/\dd z=2v_c\delta(z)$. 
The dominating energy loss process is ionisation on neutral gas in the disk, which can again be assumed to be infinitesimal~\citep{Schlickeiser1994,Evoli2019_AfterAMS}:
\begin{equation}
    \left(\frac{\dd p}{\dd t} \right)_{j,\text{ion}}=2h\dot{p}_j\delta(z)\, ,
\end{equation}
where we use the formula given by~\citet{Schlickeiser1994} and correct a typo as described in~\citet{Jacobs2022}.

There are three ways to produce CRs of species $j$, accounted for by the right hand side of eq.~\eqref{eqn:TPE}:
\begin{eqnarray}
    Q_{j}(p)&=&2h\delta(z)Q_{\text{prim},j}(p) \nonumber \\
    &+&\sum_{k>j}\frac{\psi_k}{\gamma \tau_{k\to j}}+2h\delta(z)\beta c n_{\mathrm{gas}}\sum_{k>j}\sigma_{k\to j}\psi_k\, .
    \label{eqn:sources}
\end{eqnarray}
The first term describes injection by a population of sources, where the dominating injection mechanism is believed to be diffusive shock acceleration, which in the linear strong shock case results in a $\R^{-2}$ spectrum~\citep{Malkov2001}. 
To account for non-linear modification effects~\citep[e.g.][]{Diesing2021}, here we assume $\R^{-2.3}$ for all primaries with the abundances given by~\cite{Evoli2020_Be}.
Unstable heavier elements can decay into species $j$, resulting in the second term of \eqref{eqn:sources}.
Furthermore, heavier CRs can spallate into species $j$ by interacting with the interstellar gas in the disk (the third term of eq. \eqref{eqn:sources}). 
The uncertainty on those interaction cross-sections will be a source of systematic errors, which is discussed in Sec.~\ref{sec:discussion}. 
The disk height is $\unit{100}{pc}$ and the hydrogen number density is assumed to be $n_{\mathrm{gas}}=\unit{1}{cm^{-3}}$, similar to what was adopted in~\citet{Ferriere2001,Phan2021}.

We solve the diffusion equation separately in the disk and in the halo, demanding continuity of the CR density and flux at the interface and in the disk, in addition to enforcing free escape at the halo's boundaries.
We take into account the entire nuclear chain up to iron.
At low energies, the fluxes in the solar system are altered by solar modulation. 
To account for this we use a force field model by~\cite{Gleeson1967} with a solar modulation potential of $\psi=\unit{700}{MeV}$ that we obtained by a fit to AMS-02 Oxygen data~\citep{Aguilar2021}.
We use these fluxes to calculate the $\mathrm{B/C}$ ratio, where we make the same assumptions as the AMS-02 collaboration on isotopic composition, namely pure $\mathrm{^{12}C}$ as well as a combination of $\mathrm{^{10}B}$ and $\mathrm{^{11}B}$~\citep{Aguilar2016}.
These results are fitted to AMS-02 data~\citep{Aguilar2016} and recent Voyager data~\citep{Cummings2016}. 
The latter have been measured outside the heliosphere and are therefore not subject to solar modulation.

It can be shown that for high energies the $\mathrm{B/C}$ ratio depends on an averaged diffusion ratio:
\begin{equation}
\frac{\psi_{\rm B}}{\psi_{\rm C}}\sim\frac{H-h}{\kappa_{\rm halo}}+\frac{h}{\kappa_{\mathrm{disk}}}.
\label{eq:bchigh}
\end{equation}
Since the averaged diffusion coefficient in the Galaxy cannot change significantly, as shown by~\citet{Hooper2018,Profumo2018}, the impact of a suppressed diffusion in the disk on $\mathrm{B/C}$ is rather small and is degenerate from a change of $H/\kappa_{\mathrm{halo}}$. 
To break the degeneracy between $H$ and $\kappa_{\mathrm{halo}}$, $\mathstrut^{10} \text{Be}$ data are needed, since they allow to determine the propagation time. Because a suppressed diffusion coefficient in the disk inhibits the escape of freshly produced $\mathstrut^{10} \text{Be}$ into the halo, these measurements can also potentially allow to determine $\alpha$.

For kinetic energies per nucleon below a few GeV we use available data presented in \citet{Garcia-Munoz1977,Wiedenbeck1980,Garcia-Munoz1981,Lukasiak1994,Lukasiak1999,Connell1998,Yanasak2001,Nozzoli2021}, but there are no published data at energies above $E_{k/n}\approx\, 1\,{\rm GeV/n}$. 
At this point, we will consider preliminary AMS-02 data~\citep{Wei2022}. 
In addition, we will forecast the constraining power of upcoming HELIX data~\citep{Park2019} by fitting to mock data. 
For generating these, we have adopted the model parameters as determined by fitting to all other data sets. 
The energy binning is chosen in accordance with~\citet{Park2019} and for each energy bin, we have drawn a sample from a normal distribution centred on the model and with errors as given in~\citet{Park2019}. 

In order to find the best-fit parameters we adopt a Gaussian log-likelihood,
\begin{equation}
    -2 \ln \lr{\mathcal{L}(\theta)}=\chi^2(\theta)=\sum_{d,i}\lr{\frac{\mathcal{O}_{d,i}-\mathcal{O}^{(m)}_{d,i}(\theta)}{\sigma_{d,i}}}^2,
\end{equation}
where $d$ runs over the observables $\mathrm{B/C}$ and $\mathrm{^{10}Be/^9Be}$ and $i$ runs over rigidities $\R_i$ for $\mathrm{B/C}$ AMS-02 data or kinetic energies per nucleon $E_{k/n,i}$ else.
The errors are the statistical and systematic uncertainties summed in quadrature. 
We have refrained from taking into consideration the possibility of correlated errors~\citep[see, e.g.][]{2019A&A...627A.158D} as no official information on this has been provided by the collaborations; we also consider this to be of lesser importance than, for instance, in searches for excesses well-localised in rigidity~\citep[e.g.][]{2020PhRvR...2b3022B}.  

Apart from the best-fit values, we want to quantify the allowed parameter intervals. Therefore, we have performed a Monte Carlo Markov Chain (MCMC) study which also provides robustness in the identification of the best-fit parameter values.
We make use of the \texttt{emcee} package~\citep{Foreman-Mackey2013} and use uniform priors as listed in Table~\ref{tab:priors}.
\begin{table*}
\caption{
Boundaries of the uniform priors on the model parameters. The lower bound on $\R_l$ is chosen to be smaller than the rigidity at which the diffusion timescale becomes smaller than the advection and energy loss timescales.
}
\centering
\begin{tabular}{c|c|c|c|c|c|c}
$\alpha$ & $\kappa_{\mathrm{halo}}\,[\mathrm{pc^2/yr}]$ & $H\,[\mathrm{kpc}]$ & $v_c\,[\mathrm{km/s}]$ & $\delta$ & $\delta_l$ & $\R_l\,[\mathrm{GV}]$ \\
\hline
$[0,1]$ & $[0.001,1]$ & $[0.1,20]$ & $[0.01,100]$ & $[0,2]$ & $[-4,4]$ & $[2.2,20]$\\
\end{tabular}

\label{tab:priors}
\end{table*}
For model comparison, we use a likelihood ratio test, modified to take into account the fact that the parameter $\alpha$ is bounded~\citep{Cowan2011}.


In the following we will investigate the null hypothesis, i.e. that there is no suppression, and the full model for different data sets. 
Each of them contains the AMS-02 and Voyager B/C data, together with all published $\mathrm{^{10}Be/^9Be}$ data. 
The first setup (``w/o prelim.'') reflects our best knowledge to date and only accounts for existing data. The second setup (``w/ prelim.'') additionally takes into account preliminary AMS-02 data on $\mathrm{^{10}Be/^9Be}$~\citep{Wei2022}. We expect the latter to be able to break the degeneracy of $\kappa$ and $H$. Additionally we will predict the effects of upcoming HELIX data as described above in the setup called ``w/o forecast''.

%% file: sections/Results.tex
\section{Results}
\label{sec:results}

\begin{figure*}
\centering
\includegraphics[scale=1]{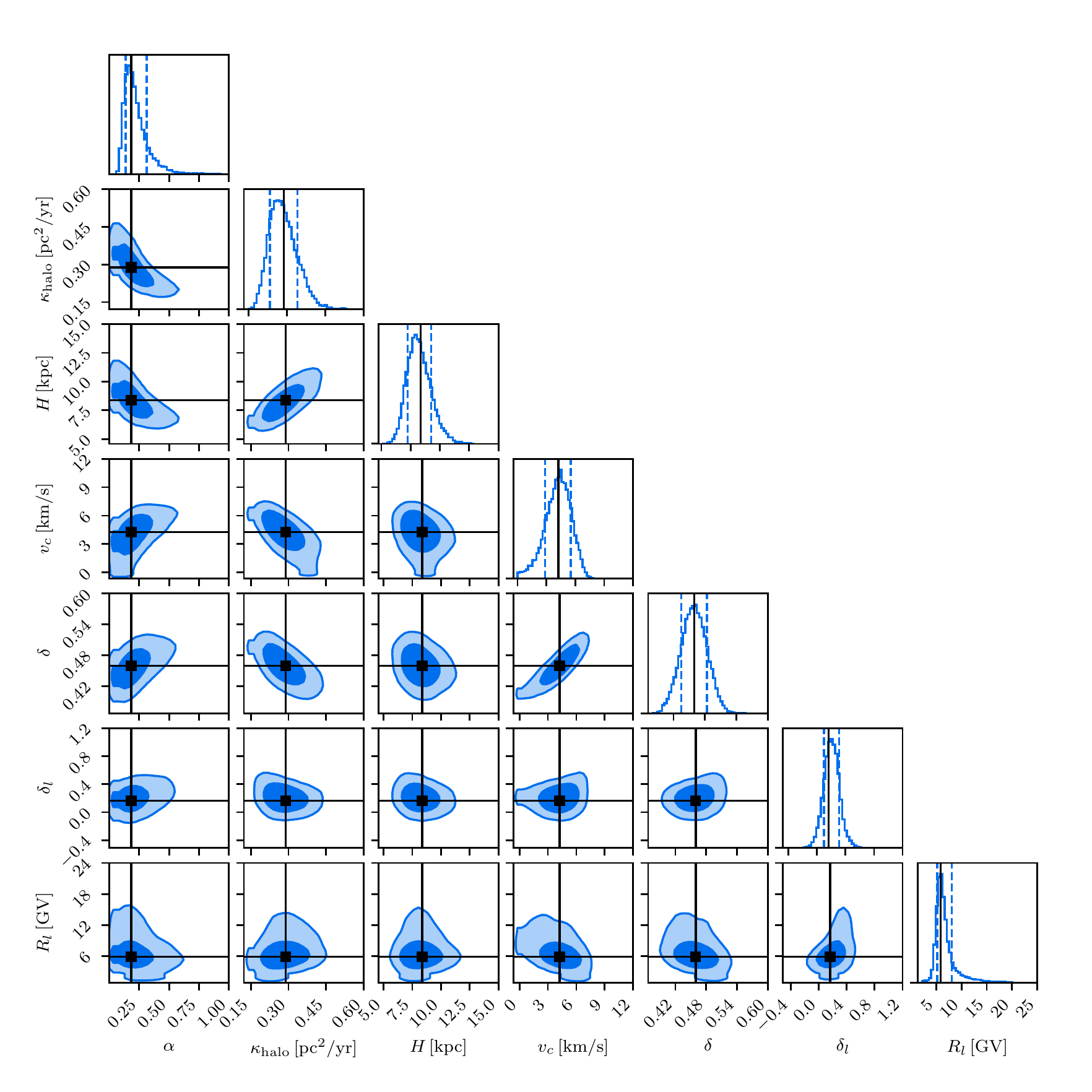}
\caption{Corner plot of the MCMC scan of the full model including preliminary data, ``w/ prelim.''. The 2D marginalised posteriors are shown on the lower triangle. The dark/bright blue areas corresponds to the $\unit{68}{\%}$/$\unit{95}{\%}$ quantiles. Similarly, the principal diagonal displays the 1D marginalised posterior with the $\unit{16}{\%}$ and $\unit{84}{\%}$ quantiles marked as blue dashed lines. The black lines indicate the best fit values as listed in Table~\ref{tab:bestfit}.}
\label{fig:corner_preliminary}
\end{figure*}
The corner plot of the MCMC for the full model with preliminary data (``full w/ prelim.'') is shown in Fig.~\ref{fig:corner_preliminary}. 
The lower left triangle displays the 2D marginalised posteriors of all parameter combinations. 
The dark blue (bright blue) regions represent the $\unit{68}{\%}$ ($\unit{95}{\%}$) quantiles. 
On the diagonal the 1D marginalised posteriors are shown with the $\unit{16}{\%}$ and $\unit{84}{\%}$ quantiles indicated by blue dashed lines. 
Both in the 1D and 2D posteriors the best fit values are marked with solid black lines, see also Table~\ref{tab:bestfit}, where we give the $\chi^2$ of the setups as well. 
The first column, which shows the posterior as a function of $\alpha$ clearly indicates that the null hypothesis is firmly ruled out.
An anticorrelation between $\kappa_{\mathrm{halo}}$ and $\alpha$ exists, see eq. \eqref{eq:bchigh}, where a smaller $\alpha$ partially offsets the effects of a larger $\kappa_{\mathrm{halo}}$. 
The degeneracy of $\kappa_{\mathrm{halo}}$ and halo height $H$ is largely broken and the halo height can be constrained to $H=\ut{8.2}{1.1}{0.9}\,\mathrm{kpc}$, however some correlation can still be observed.
Another degeneracy exists between $\delta$ and $v_c$. 
This is a well-documented effect observed in advection-diffusion models, see e.g. Fig.~2 of~\citet{2010A&A...516A..66P}. 
It is due to the fact that in order to match the secondary-to-primary ratio over a limited rigidity range, a larger advection speed $v_c$ can be compensated by a larger $\delta$, that is a stronger rigidity dependence of $\kappa$. 
The best-fit values are generally in good agreement with the median of the distribution, but for the low energy break rigidity volume effects in the case $\delta=\delta_l$ lead to an extended non-Gaussian tail.
The corner plots of the other setups and the null hypothesis are given and explained in appendix~\ref{sec:corner}.

\begin{figure}
\includegraphics[scale=1]{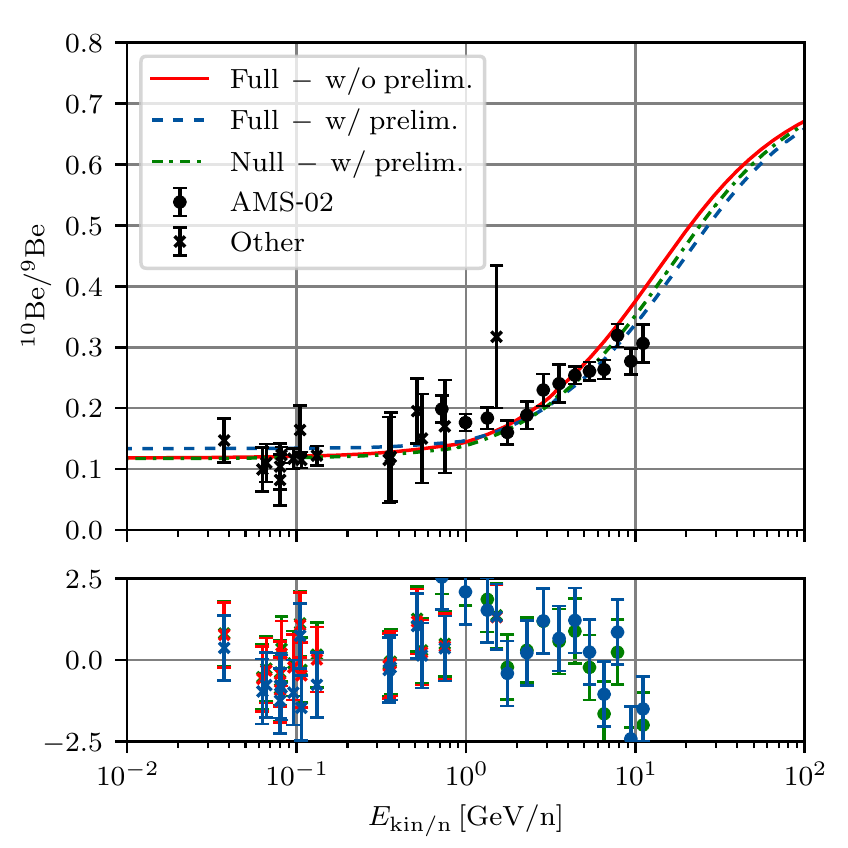}
\caption{Best-fit spectra for the Beryllium ratio to the setups with only existing data (red solid), preliminary AMS-02 data (blue dashed) and the corresponding null hypothesis $\alpha=1$ (green dot-dashed). The null hypothesis predicts a lower Be ratio at low energies, and a transition to the high energy limit earlier, since the best-fit halo height is smaller (for more details see text). The lower
panel is the pull plot.}
\label{fig:Be_spectra}
\end{figure}
Figure \ref{fig:Be_spectra} shows the best-fit $\mathrm{^{10}Be/^9Be}$ spectra as a function of kinetic energy per nucleon for the full model with and without preliminary data (blue dashed and red solid lines) and the null hypothesis with preliminary data (green dot-dashed lines). 
At low energies the ratio is determined by an interplay between diffusion and advection. 
Since Be is purely secondary, it is exclusively produced in the disk. 
A faster transport in the disk by advection or diffusion leads to freshly spallated $\mathrm{^{10}Be}$ being transported out of the disk into the halo, where it will decay. 
Since the solar system is located in the disk, this results in a lower observed ratio. 
Therefore, the Be ratio at low energies is especially sensitive to the suppression of the diffusion coefficient, which has little effect in the propagation afterwards. 
The high energy limit, where $\mathrm{^{10}Be}$ can be considered as stable due to time dilatation, is given by the ratio of the production cross sections~\citep{Maurin2022} and is dominated by their uncertainties. 
The energy at which the transition happens is determined by the energy as which the decay time of particles becomes larger than the escape time. 
Furthermore, the steepness of the rise is determined by the escape time.
Together the above effects explain, why our setup can constrain $\alpha,\ H$ and $\kappa_{\mathrm{halo}}$, as shown in Fig.~\ref{fig:corner_preliminary}.
The decrease in $\alpha$ together with the increase in $H$ of the full model ``w/ preliminary'' compared to the null hypothesis results in a flatter transition to the high energy limit as preferred by the AMS-02 data. 
Without the AMS-02 data the transition is unconstrained, resulting in both $\alpha,\kappa$ and $H$ being essentially unconstrained, as visible in Fig. \ref{fig:corner_nopre}.

\begin{figure}
\includegraphics[scale=1]{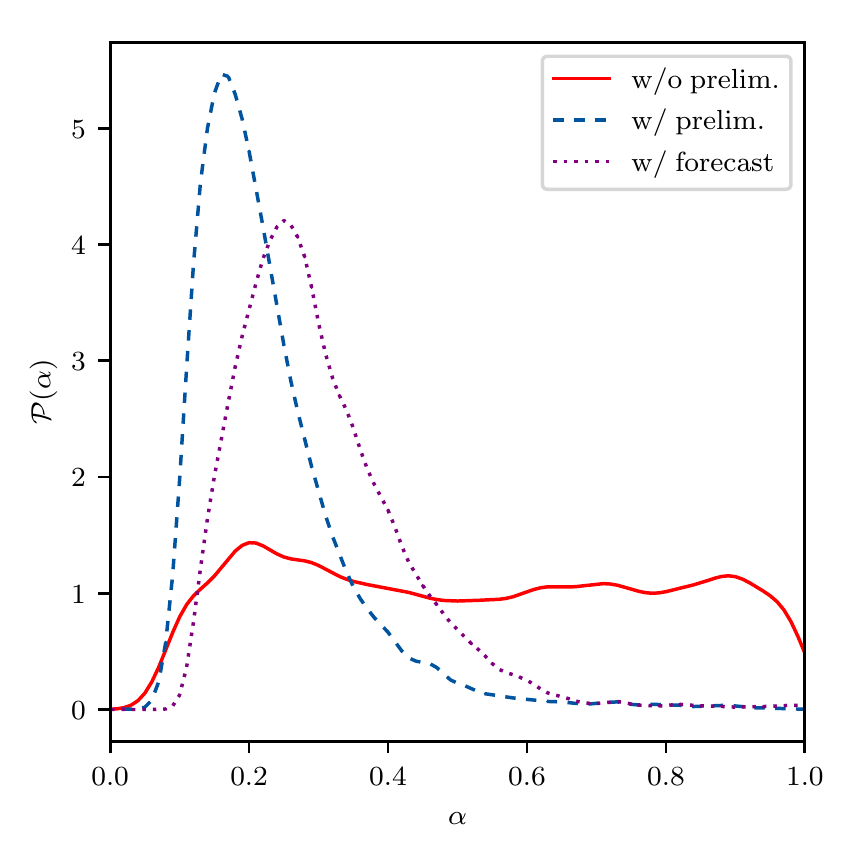}
\caption{Marginalised posterior of the scaling factor $\alpha$. The setup with existing data only (red solid) shows no preference for a suppression. The setup with preliminary AMS-02 data, ``w/ prelim.'', allows constraining the scaling factor to around $\ut{0.20}{0.10}{0.06}$, with a model preference of $\unit{3.5}{\sigma}$. Upcoming HELIX data will be able to constrain the ``w/ prelim.'' best fit model to $\unit{6.8}{\sigma}$. Due different effects the increase in the significance difficult to estimate from the posterior (see text for details).}
\label{fig:1D_posterior_alpha}
\end{figure}
The 1D marginal posterior as a function of $\alpha$ for all three data sets is shown in Fig.~\ref{fig:1D_posterior_alpha}. 
Here, we have used a Gaussian kernel density estimation with a width of $0.1$. 
The posterior for existing data is indicated by the red solid line, the one including preliminary data by the blue dashed line and for the forecast by the dotted purple line. 

We give the median and estimate its error by its distance to the $16\,\%$ and $84\,\%$ quantiles in Table \ref{tab:bestfit}. 
For the setup ``w/o prelim.'' the posterior is flat and extends to $\alpha=1$. 
The observed decline towards $\alpha=1$ is a result of the chosen prior $\alpha<1$ and the convolution with the Gaussian kernel. 
Values larger than $1$ are unconstrained by our model, since the residence time in the disk would become negligible. 
Thus without preliminary AMS-02 data, we do not find a preference for a suppressed diffusion coefficient. 
It comes as no surprise that $H$ and $\kappa$ are also still degenerate.  
Including preliminary AMS-02 data, allows to constrain the median to $\alpha=\ut{0.20}{0.10}{0.06}$. 
In order to compare models we use a likelihood ratio test and account for the null hypothesis being at the boundary of our parameter space~\citep{Cowan2011}.
This results in a $\unit{3.5}{\sigma}$ preference for a model with suppressed diffusion. 
Should the trend continue with upcoming HELIX data, we forecast a detection by $\unit{6.8}{\sigma}$. 
This increase of significance is in contrast to the naive expectation gleaned from looking at the posteriors. 
However, there are several reasons why we deem the modified test statistics to be more reliable. 
First, the MCMC scan lacks statistics so far in the tails of the posterior and secondly, we expect volume effects to contribute in the tail as well. 
Hence, the median and its errors are well estimated by the posterior, but the model significance has to be determined using test statistics.


\begin{table*}
\centering
\caption{
Best-fit and median of the setups, the errors are estimated as the distance to the $\unit{16}{\%}$ and $\unit{84}{\%}$ quantiles of the marginalised 1D posteriors respectively. 
}
\setlength{\tabcolsep}{10pt} 
\renewcommand{\arraystretch}{1.25} 
\begin{tabular}{c|c c|c c|c c}
\setlength{\tabcolsep}{1000pt}
 & \multicolumn{2}{c|}{Full - w/o prelim. } & \multicolumn{2}{c|}{Full - w/ prelim.} & \multicolumn{2}{c}{Null - w/ prelim.}\\
 & median & best-fit & median & best-fit & median & best-fit \\ 
\hline
$\alpha$ & $\ut{0.5}{0.3}{0.3}$ & $0.8$& $\ut{0.20}{0.10}{0.06}$ & $0.19$ &  -   & - \\
$\kappa_{\mathrm{halo}}\,[\text{pc}^2/\text{yr}]$ & $\ut{0.21}{0.11}{0.05}$ & $0.17$& $\ut{0.28}{0.06}{0.05}$ & $0.29$& $\ut{0.174}{0.009}{0.009}$   & $0.172$\\
$H\,[\text{kpc}]$ & $\ut{6.3}{3.5}{1.5}$ & $5.1$& $\ut{8.2}{1.1}{0.9}$ & $8.2$ & $\ut{6.4}{0.4}{0.4}$  & $6.2$\\
$v_c\,[\text{km}/\text{s}]$ & $\ut{4.3}{1.3}{1.5}$ & $4.3$ & $\ut{4.2}{1.2}{1.3}$ & $4.2$& $\ut{6.9}{0.6}{0.7}$  & $6.5$ \\
$\delta$ & $\ut{0.46}{0.03}{0.02}$ & $0.46$ & $\ut{0.46}{0.02}{0.02}$ & $0.45$ & $\ut{0.511}{0.017}{0.016}$  & $0.506$ \\
$\delta_l$ & $\ut{0.24}{0.12}{0.11}$ & $0.21$ & $\ut{0.2}{0.1}{0.1}$ & $0.16$ & $\ut{0.5}{0.29}{0.19}$ & $0.35$\\
$R_l\,[\text{GV}]$ & $\ut{6.5}{2.6}{1.2}$ & $6.2$ & $\ut{6.1}{1.9}{0.9}$ & $5.8$ & $\ut{5.0}{7.0}{3.0}$ & $6.0$\\ 
$\chi^2$ & - & $67.7$ & - & $99.6$ & -   & $111.9$ \\
$\text{dof}$ & $85$ & $85$ & $98$ & $98$ & $99$ & $99$ \\
\end{tabular}

\label{tab:bestfit}
\end{table*}

%% file: sections/Discussion.tex
\section{Discussion}
\label{sec:discussion}
We have seen in the previous section that preliminary data on Beryllium isotopes from the AMS-02 experiment favour a smaller diffusion coefficient in the disk than in the halo. 
As far as the interpretation is concerned, our analysis does not allow distinguishing two scenarios for the origin of this suppressed diffusion: 
Either the diffusion coefficient is suppressed (more or less) everywhere in the disk, for instance due to different turbulence generation or damping than in the halo; or the diffusion coefficient is strongly suppressed in localised regions of a certain filling fraction.


For the scenario where the suppression is due to bubbles of low diffusivity, we have estimated $\alpha$ for a given ratio $(\kappa_{\mathrm{low}}/\kappa_{\mathrm{high}})$ by numerically simulating the transport of a large number of CRs with stochastic differential equations (SDEs). 
To this end, we have modelled the disk as a periodic medium with a diffusion coefficient $\kappa_{\mathrm{high}}$ and spherical regions of suppressed diffusion coefficient $\kappa_{\text{low}}$, arranged on a cubic grid. 
For the ratio of $(\kappa_{\mathrm{low}}/\kappa_{\mathrm{high}})$ we have adopted the values $10^{-3}$ or $10^{-2}$. 
The sources of CRs have been assumed to be uniformly distributed in the medium. 
We have computed the coarse-grained diffusion coefficient as $\kappa_{\mathrm{disk}}$ from the mean-square displacement of particles,
\begin{equation}
\kappa_{\mathrm{disk}} = \lim_{t \to \infty} \frac{1}{2} \frac{\dd \langle (\Delta r)^2 \rangle}{\dd t} \, .
\end{equation}
We have checked that the exact configuration (e.g. simple cubic, body-centred cubic or face-centred cubic) does not matter for the asymptotic value of $\kappa_{\mathrm{disk}}$, but only the filling fraction. 

\begin{figure}
    \centering
    \includegraphics[scale=1]{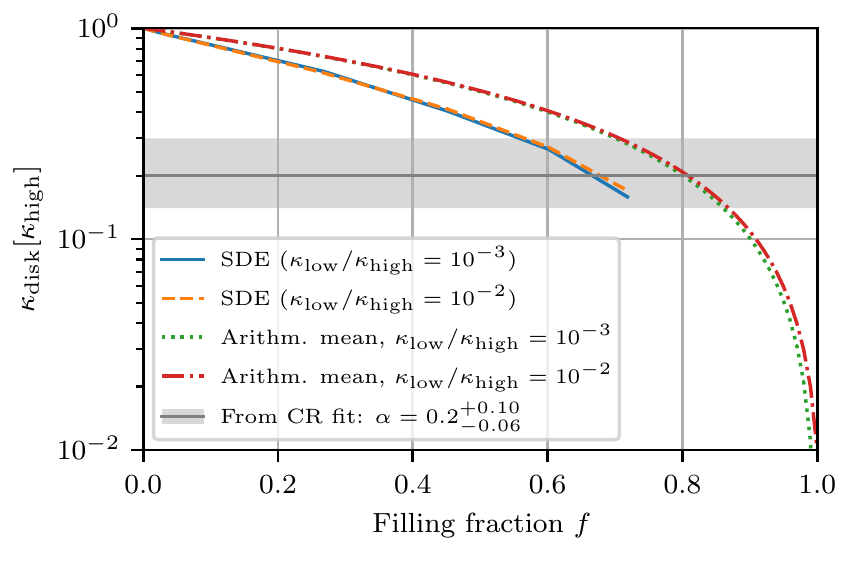}
    \caption{Effective diffusion coefficient $\kappa_{\mathrm{disk}}$ from the SDE computation as a function of the volume filling fraction $f$. The solid blue (dashed orange) line shows $\kappa_{\mathrm{disk}}$ in unit of $\kappa_{\text{high}}$, assuming a ratio $\kappa_{\text{low}}/\kappa_{\text{high}}$ of $10^{-3}$ ($10^{-2}$). The sources of CRs are assumed to be distributed uniformly in the disk. Note that the lines end at the maximum possible filling fraction of $\sim 74 \, \%$ for spherical inclusions. For comparison, the green dotted (red dot-dashed) line shows the arithmetic mean $(\kappa_{\mathrm{low}} / \kappa_{\mathrm{high}}) f + (1-f)$ assuming $\kappa_{\text{low}}/\kappa_{\text{high}}$ of $10^{-2}$ ($10^{-3}$). The grey line and band highlight the suppression found from the CR fit, $\alpha = 0.2^{+0.10}_{-0.06}$.}
    \label{fig:filling_fraction_suppression}
\end{figure}
In Fig.~\ref{fig:filling_fraction_suppression}, we show the suppression $\alpha$ achieved as a function of the volume filling fraction $f$ of bubbles of low diffusion in the disk. 
The value $\alpha = 0.2^{+0.10}_{-0.06}$, found in the CR parameter study can be realised for a filling fraction of $\sim 66 \, \%$. 
We also compare to the arithmetic mean achieved for a given filling fraction.

In the literature, the value typically estimated for the filling fraction $f$ is lower,
\begin{align}
    f &\equiv \frac{N_{\mathrm{bubbles}} V_{\mathrm{bubble}}}{V_{\mathrm{disk}}} = \frac{4}{3} \frac{\mathcal{R}_{\text{bubble}} \tau_{\text{bubble}} R_{\text{bubble}}^3}{R_{\mathrm{disk}}^2 h} \nonumber \\
    &= 0.0048 \left( \frac{\mathcal{R}_{\text{bubble}}}{0.03 \, \text{yr}^{-1}} \right) \left( \frac{\tau_{\text{bubble}}}{10^5 \, \text{yr}} \right) \left( \frac{R_{\text{bubble}}}{0.03 \, \text{kpc}} \right)^3 \nonumber \\
    & \quad \times \left( \frac{R_{\text{disk}}}{15 \, \text{kpc}} \right)^{-2} \left( \frac{h}{0.1 \, \text{kpc}} \right)^{-1} \, ,
\end{align}
where $N_{\mathrm{bubbles}}$ is the number of bubbles, $\mathcal{R}_{\text{bubble}}$ their rate, $R_{\text{bubble}}$, $V_{\mathrm{bubble}}$ and $\tau_{\text{bubble}}$ are the radius, volume and lifetime of one bubble and $R_{\text{disk}}$, $V_{\text{disk}}$ denote the radius and volume of the Galactic disk. 
In fact, a filling fraction of $\sim 60 \, \%$, which would be in agreement with the determined value of $\alpha$, would require a bubble radius of $150 \, \text{pc}$. 

We note that such a high filling fraction does not contradict H.E.S.S. observations~\citep{Hooper2017}, as long as the sources of electrons are distributed in both high and low diffusion zones. 
Hence, zones of low diffusivity in our Galaxy might be more common and extended than previously assumed~\citep{Manconi2020,Profumo2018,Johannesson2019}. 
Interestingly, recent studies of TeV gamma ray halos around pulsars allow for similar sizes of $\unit{100}{pc}$ of the low diffusion region~\citep{Schroer2023}.

As discussed in Sec.~\ref{sec:results}, the reason that the data prefer a suppressed diffusion coefficient in the disk is that the transition from the low to the high energy limit in the observed $\mathstrut^{10} \mathrm{Be} / \mathstrut^{9} \mathrm{Be}$ is relatively flat. 
Since the~\citet{Evoli2018} cross-sections used here result in a relatively high $\mathstrut^{10} \mathrm{Be} / \mathstrut^{9} \mathrm{Be}$, other cross-section parameterisations will likely lead to a smaller filling fraction, resulting also in a smaller bubble size.

It is worth mentioning that the filling fraction of low diffusion zones fitted here differs from previous analyses~\citep{Johannesson2019}. The main reason is the inclusion of Be data in our analysis, which allows us to constrain the diffusion coefficient in the disk and estimate the filling fraction of low diffusion zones. 
Furthermore, we adopt a different coarse-graining procedure to relate the diffusion coefficient in the disk $\kappa_{\rm disk}$ with the filling fraction of low diffusion zones $f$. 
Previous analyses \citep[e.g.][]{Johannesson2019} assume that particles start in the centre of a low diffusion zone, which is surrounded by a high-diffusion zone and then use the characteristic diffusion times to calculate an \textit{effective} diffusion coefficient in the disk used for CR propagation afterwards. 
This procedure could lead to $\kappa_{\rm disk}$ being a few times suppressed with respect to $\kappa_{\rm halo}$ even for a filling fraction of a few per cent only. 
However, this approach ignores the fact that particles, after leaving the high diffusion zone and propagating in the halo, could come back and cross the disk many times before finally escaping the Galaxy. 
When particles reenter the disk, they have a lower than one probability to travel in other low diffusion zones and this could lead to $\kappa_{\rm disk}$ larger than previously expected for the same filling fraction (see also Fig. \ref{fig:filling_fraction_suppression} and the discussion above).

 As a remark, we would like to stress also that our results have been obtained using a linear model where the diffusion coefficients are determined by a fit to CR data. There exists also non-linear Galactic CR propagation models in which the diffusion coefficient is determined by turbulence generated via CRs themselves. Many of these models, however, predict $\kappa_{\rm disk} > \kappa_{\rm halo}$ at least for certain regions in the halo which seems to contrast with the results in our linear model. 
 For example, \citet{recchia2016} found that balancing the growth rate of CR-induced turbulence and the non-linear Landau damping rate (see e.g. \citealt{ptuskin2003} for more details) results in the diffusion coefficient being highest in the disk and decreasing linearly with $|z|$ up to $|z|=H$. 
 Recently, \citet{Evoli2018} revisited this type of non-linear model without imposing the free-escape boundary condition at $|z|=H$ to study the origin of the halo. 
 The authors took into account also other types of turbulence-damping mechanisms and, more importantly, the injection of turbulence due to supernova explosions. 
 In this case, the diffusion coefficient is also relatively high in the disk and reaches a minimum around $|z|\simeq 1$ kpc but increases again at larger $|z|$. 
 We note however that these models do not consider the geometry of the Galactic magnetic field which could be important in modelling CR-induced turbulence \footnote{This is because turbulence is typically treated in the form of Alfv\'enic plasma waves propagating along the large-scale ordered Galactic magnetic field.}. 
 Given that a suppressed diffusion coefficient in the disk is preferred by preliminary AMS-02 data, it would be interesting to revisit these non-linear models with a more precise geometry of the Galactic magnetic field and confront them with unstable secondary CR data.

%% file: sections/SummaryConclusion.tex
\section{Conclusion}
We have presented an extended model of Galactic CR transport, taking into account the presence of regions of suppressed diffusivity in the Galactic disk.
To this end, we have solved the CR transport equation assuming two diffusion coefficients, one in the disk ($\kappa_{\rm disk}$) and one in the halo ($\kappa_{\rm halo}$). 
Fitting this model to \emph{existing} B/C and $\mathrm{^{10}Be/^9Be}$ data by AMS-02, Voyager and several other experiments, we have found no preference for a suppressed diffusion coefficient in the disk. This has to be expected since these datasets do not even break the degeneracy between halo height and diffusion coefficient. 
Adding \emph{preliminary} AMS-02 data on $\mathrm{^{10}Be/^9Be}$ reveals a $\unit{3.5}{\sigma}$ preference for a suppressed diffusion coefficient in the disk with a best-fit value of $\alpha = \kappa_{\rm disk}/\kappa_{\rm halo} = \ut{0.20}{0.10}{0.06} $. 
Additionally, these data constrain the halo height to $H=\ut{8.9}{1.3}{1.0}\,\mathrm{kpc}$ and the diffusion coefficient in the halo to $\kappa_{\rm halo} = \ut{0.28}{0.06}{0.05}\,\mathrm{pc^2/yr}$. 
We predict that, if the preliminary findings by AMS-02 are confirmed, upcoming HELIX data will be able to detect a suppressed diffusion coefficient in the disk with $\unit{6.8}{\sigma}$ significance.

The suppressed diffusion coefficient in the disk found in this work can be interpreted as the collective effect of low diffusion zones present inside the Galactic disk. 
In fact, these low diffusion zones have been inferred from gamma-ray observations of regions surrounding supernova remnants or PWNe~\citep[e.g.][]{Gabici2010,Abeysekara2017} and also inside dense molecular clouds \citep[e.g.][]{Yang2023}. 
Having found the scaling factor $\alpha=\kappa_{\rm disk}/\kappa_{\rm halo}$, we proceed to constrain the filling fraction of low diffusion zones $f$. 
The relation between $\kappa_{\rm disk}$ and $f$ is found using simulations of CR transport in an inhomogeneous diffusion medium. 
Within a given medium with low diffusion zones of filling fraction $f$, the {\it effective} diffusion coefficient of CRs is as presented in Figure~\ref{fig:filling_fraction_suppression}. 
This allows us to constrain the filling fraction to be around $66\%$, i.e.\ about 2/3 of the disk should be filled with low diffusion zones. 
Our results suggest that low diffusion zones surrounding potential CR sources such as supernova remnants and pulsars might be more ubiquitous than previously expected.

%% file: sections/Appendix.tex
\section{Two Zone Diffusion Model}
The transport equation for CRs in steady state reads:
\begin{eqnarray}
        \frac{\partial }{\partial z}\lr{v_c\psi_j-\kappa(z,\R)\frac{\partial \psi_j}{\partial z}} +\frac{\partial}{\partial p}\left[\lr{\frac{\dd p}{\dd t}}\psi_j-\frac{p}{3}\frac{dv_c}{dz} \psi_j\right]  
        +\frac{1}{\gamma \tau_j}\psi_j + 2h\delta(z)\beta c n_{gas}\sigma_j \psi_j = Q_{j}(p) \, .
    \label{eq:tpeap}
\end{eqnarray}
In a two zone model, this equation is solved using the following steps:
\begin{enumerate}
    \item The transport equation is solved in the halo ($h < |z| < H$) with the boundary condition $\psi(H)=0$.
    \item We solve the equation in the disk and demand that both the density $\psi$ and its spatial flux be conserved at $|z| = h$.
    \item We integrate the transport equation over an infinite volume around $z=0$.
\end{enumerate}
In the following we will denote all properties in the halo with a $o$ for outer and in the disk with $i$ for inner.\newline
Starting at (i),  the transport equation at $z>0$ reads:
\begin{eqnarray}
        &&\frac{\partial }{\partial z}\lr{v_c\psi_j-\kappa_o\frac{\partial \psi_j}{\partial z}}+\frac{1}{\gamma \tau_j}\psi_j = 0 \,, \qquad |z|>h, 
        \label{eq:tpezl0}\\
        && \frac{\partial }{\partial z}\lr{v_c\psi_j-\kappa_i\frac{\partial \psi_j}{\partial z}}+\frac{1}{\gamma \tau_j}\psi_j = 0 \,, \qquad 0<|z|<h, \label{eq:tpezl0-2}
\end{eqnarray}
where we have neglected decay into species $j$. \newline
Now we make the Ansatz $\psi_j=C\ex{\lambda z}$ and introduce the decay rate $\Gamma_j=\lr{\gamma \tau_j}^{-1}$. Plugged into eq. \eqref{eq:tpezl0} this results in:
\begin{eqnarray}
    \psi_j = \ex{\frac{zv_c}{2\kappa_o}}\lr{A_o\ex{\frac{z}{\Tilde{z}_o}}+B_o\ex{-\frac{z}{\Tilde{z}_o}}}\mathrm{with} \quad \frac{1}{\Tilde{z}_o}=\sqrt{\lr{\frac{v_c}{2\kappa_o}}^2+\frac{\Gamma}{\kappa_o}} .
\end{eqnarray}
Using the boundary condition $\psi(z=H)=0$ gives:
\begin{eqnarray}
    A_o=-B_o\ex{-\frac{2 H}{\Tilde{z}_o}}.
\end{eqnarray}
Now we can move to step (ii). The solution for eq. \eqref{eq:tpezl0-2} can be written in a manner analogous to eq. \eqref{eq:tpezl0} as:
\begin{eqnarray}
    \psi_j = \ex{\frac{zv_c}{2\kappa_i}}\lr{A_i\ex{\frac{z}{\Tilde{z}_i}}+B_i\ex{-\frac{z}{\Tilde{z}_i}}} \mathrm{with} \quad \frac{1}{\Tilde{z}_i}=\sqrt{\lr{\frac{v_c}{2\kappa_i}}^2+\frac{\Gamma}{\kappa_i}} .
\end{eqnarray}
Now we match the solutions at $z=h$:
\begin{eqnarray}
    \lim_{\epsilon\to 0} \left[\psi_j(h+\epsilon)-\psi_j(h-\epsilon)\right]=0
\end{eqnarray}
resulting in:
\begin{eqnarray}
    && B_o\left[-\ex{\frac{h-2H}{\Tilde{z}_o}}+\ex{-\frac{h}{\Tilde{z}_o}}\right]\ex{\frac{hv_c}{2\kappa_o}} 
    =\left[A_i\ex{\frac{h}{\Tilde{z}_i}}+B_i\ex{-\frac{h}{\Tilde{z}_i}}\right]\ex{\frac{hv_c}{2\kappa_i}}\nonumber\\
    \Rightarrow && 
        B_o=\frac{A_i\ex{\frac{h}{\Tilde{z}_i}}+B_i\ex{-\frac{h}{\Tilde{z}_i}}} {-\ex{\frac{h-2H}{\Tilde{z}_o}}+\ex{-\frac{h}{\Tilde{z}_o}}} \ex{\frac{hv_c}{2\kappa_i}-\frac{hv_c}{2\kappa_o}}.
\end{eqnarray}
If we use the conservation of the flux we get:
\begin{eqnarray}
    \lim_{\epsilon\to 0} \left[\lr{v_c\psi_j(h+\epsilon)-\kappa_o\left.\frac{\partial \psi_j}{\partial z}\right|_{h+\epsilon}}-\lr{v_c\psi_j(h-\epsilon)-\kappa_i\left.\frac{\partial \psi_j}{\partial z}\right|_{h-\epsilon}}\right]=0,
\end{eqnarray}
which can be solved for $B_o$ to give:
\begin{eqnarray}
    B_o=\frac{\ex{\frac{hv_c}{2\kappa_i}-\frac{hv}{2\kappa_o}} \left[ \lr{\frac{\kappa_i}{\Tilde{z}_i}+\frac{v_c}{2}}\ex{\frac{h}{\Tilde{z}_i}}A_i+\lr{\frac{v_c}{2}-\frac{\kappa_i}{\Tilde{z}_i}}\ex{-\frac{h}{\Tilde{z}_i}}B_i  \right]}  {\frac{v_c}{2}\left[\ex{-\frac{h}{\Tilde{z}_o}}-\ex{\frac{h-2H}{\Tilde{z}_o}}\right]-\frac{\kappa_o}{\Tilde{z}_o}\left[\ex{\frac{h-2H}{\Tilde{z}_o}}+\ex{-\frac{h}{\Tilde{z}_o}}\right]}.
\end{eqnarray}
If we now eliminate $B_o$ and express $A_i$ in terms of $B_i$:
\begin{eqnarray}
    A_i=-B_i\ex{-\frac{2h}{\Tilde{z}_i}}\lr{\frac{2\frac{\kappa_i}{\Tilde{z}_i}}{\frac{\kappa_o}{\Tilde{z}_o}\mathrm{tanh}^{-1}\lr{\frac{h-H}{\Tilde{z}_o}}-\frac{\kappa_i}{\Tilde{z}_i}}+1} \equiv B_i\frac{\gamma_2}{\gamma_1}. 
\end{eqnarray}
In order to find $B_i$ we need to integrate eq. \eqref{eq:tpeap} over an infinite volume around $z=0$.\newline
Denoting $\psi_j(z=0)\equiv\psi_{j,0}$, this gives:
\begin{eqnarray}
2v_c\psi_{j,0}-\kappa_i\left.\frac{\partial}{\partial z}\psi_j \right\vert_{0-}^{0+}+2h\beta c n_{\mathrm{gas}}\sigma_j\psi_{j,0}+\frac{\partial}{\partial p}\lr{2h\dot{p}-\frac{p}{3}\frac{\dd v_c}{\dd z}}\psi_{j,0}
=2hQ_{\mathrm{prim,j}}(p)+2h\beta c n_{\mathrm{gas}}\sum_{k>j}\sigma_{k\to j}\psi_k,
\end{eqnarray}
where the derivative can be computed for the solution for $z>0$ and the advection velocity is changing sign at the disk.\newline
Then the above equation reads:
\begin{eqnarray}
2v_c\psi_{j,0}-2\kappa_i\left.\frac{\partial}{\partial z}\psi_j \right\vert_{0}+2h\beta c n_{\mathrm{gas}}\sigma_j\psi_{j,0}+\frac{\partial}{\partial p}\lr{2h\dot{p}-\frac{2pv_c}{3}}\psi_{j,0}
=2hQ_{\mathrm{prim,j}}(p)+2h\beta c n_{\mathrm{gas}}\sum_{k>j}\sigma_{k\to j}\psi_k,
\end{eqnarray}
Now we use this equation to calculate $B_i$.
The differential equation that has to be solved is:
 \begin{eqnarray}
     \left[\lr{2v_c+2h\beta c n_{\mathrm{gas}}\sigma_j}\lr{\frac{\gamma_2}{\gamma_1}+1}-2\gamma_i\lr{\frac{\gamma_2}{\gamma_1}-1}+\frac{\partial}{\partial p}\lr{2h\dot{p}-\frac{2pv_c}{3}}\lr{\frac{\gamma_2}{\gamma_1}+1}\right]B_i=2hQ_{\mathrm{prim,j}}(p)+2h\beta c n_{\mathrm{gas}}\sum_{k>j}\sigma_{k\to j}\psi_k,\nonumber\\
 \end{eqnarray}
This is done numerically using Scipy's solve$\_$ivp function~\citep{Virtanen2020}.

\section{Additional Plots}
\label{sec:corner}
\begin{figure*}
\centering
\includegraphics[scale=1]{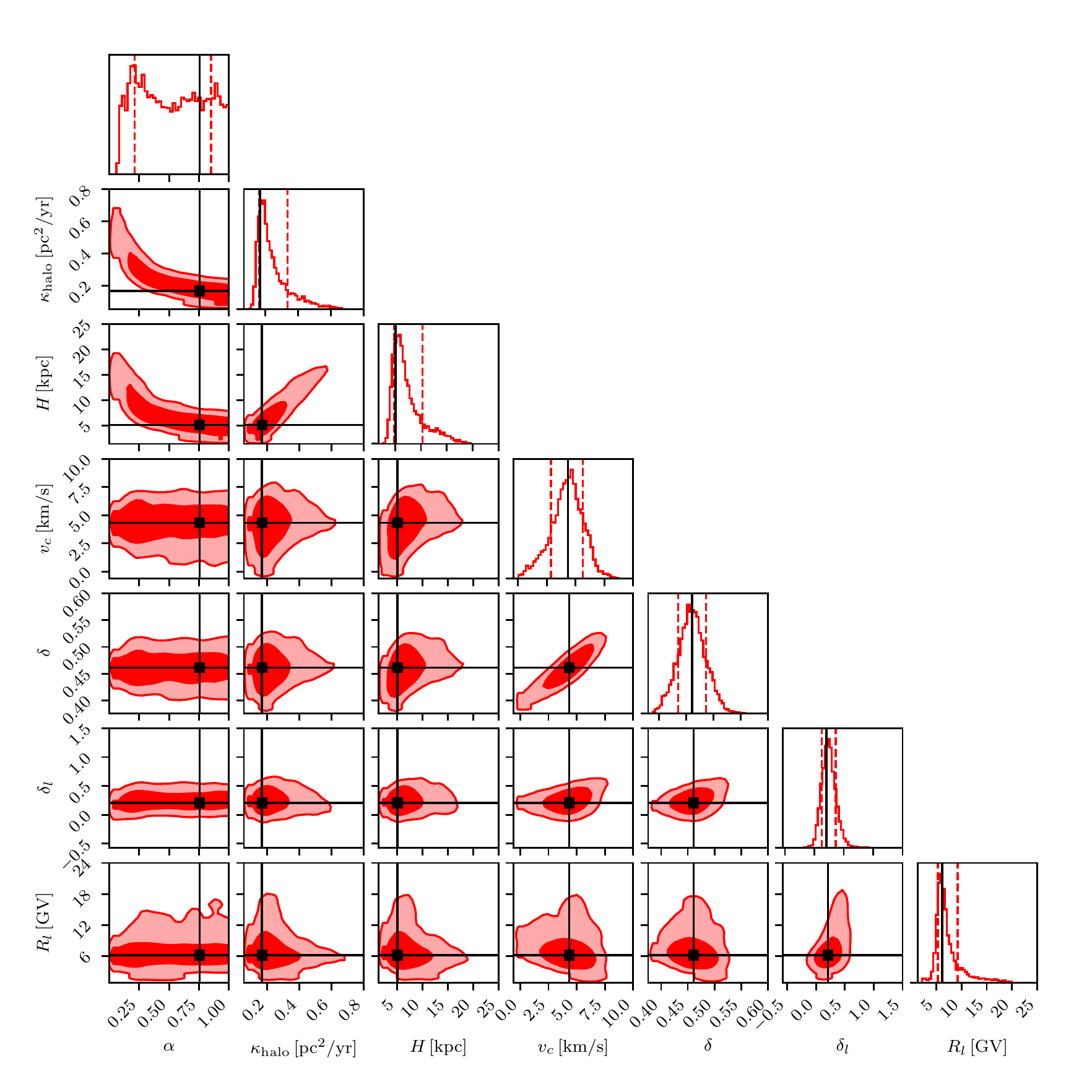}
\caption{
Same as Fig.~\ref{fig:corner_preliminary}, but for the full model excluding preliminary data, called ``w/o prelim.''
The extension of the posterior to $\alpha=1$ indicates that the data cannot constrain $\alpha$. Furthermore, the ratio of $H/\kappa_o$ is largely unconstrained.}
\label{fig:corner_nopre}
\end{figure*}

\begin{figure*}
\centering
\includegraphics[scale=1]{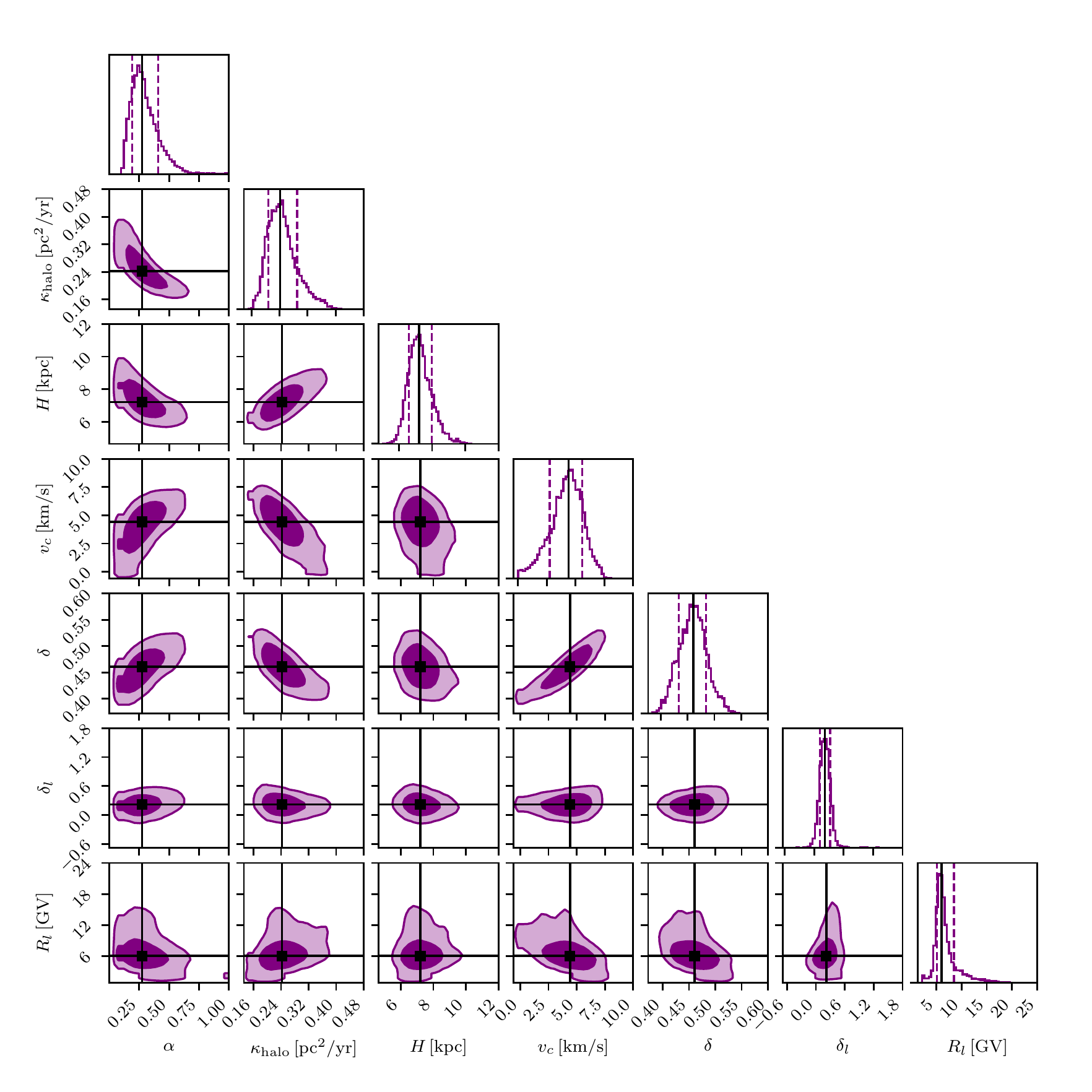}
\caption{
Same as Fig.~\ref{fig:corner_preliminary}, but for the full model including a forecast of upcoming HELIX data.}
\label{fig:corner_fore}
\end{figure*}

The corner plots of the MCMC scans for the ``Full w/o prelim.'' setup is shown in Fig.~\ref{fig:corner_nopre}. Due to the lack of Be data in the GeV region, the degeneracy between $H$ and $\kappa_o$ coming from B/C data cannot be broken. This is visible in the second column, third row. As a consequence, the suppression factor $\alpha$ is unconstrained as well. Due to the posterior being non-Gaussian the marginalised posteriors are subject to volume effects. Including both preliminary AMS-02 and a forecast for HELIX data, ``w/ forecast'', results in the corner plot shown in fig. \ref{fig:corner_fore}. Here the degeneracy between $H$ and $\kappa_o$ is broken and $\alpha$ can be constrained, as explained in Sec.~\ref{sec:discussion}.

\begin{figure*}
\centering
\includegraphics[scale=1]{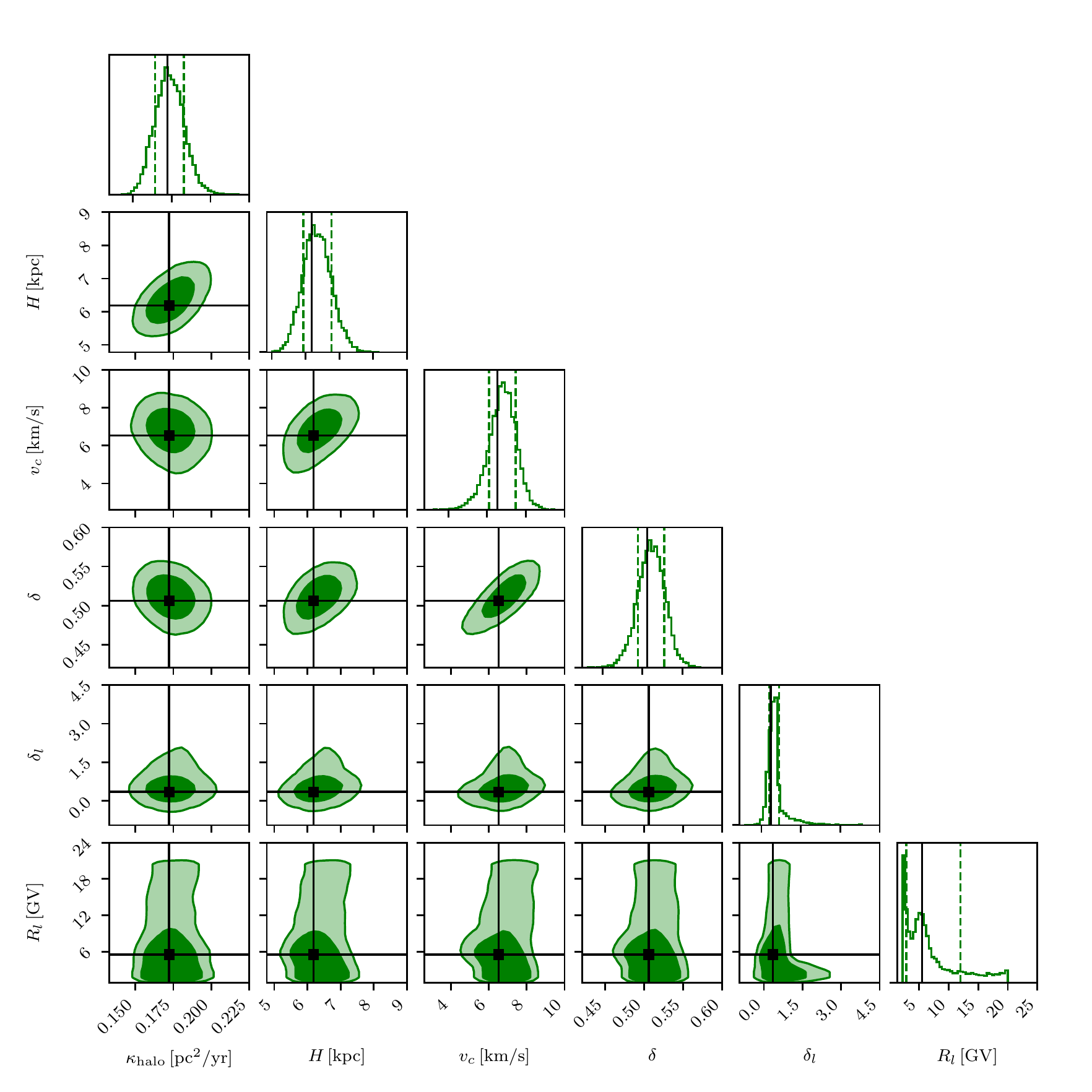}
\caption{
Same as Fig.~\ref{fig:corner_preliminary}, but for the null model including preliminary AMS-02 data.}
\label{fig:ncorner}
\end{figure*}

For the null hypothesis with preliminary data, ``Null w/ prelim.'', the corner plot is given in Fig.~\ref{fig:ncorner}. Here, the degeneracy between $H$ and $\kappa_o$ is broken again. Compared to the full model, smaller values for both parameters are preferred. In the case $\delta\approx\delta_l$ the lower break rigidity $R_l$ becomes unconstrained.

\begin{figure*}
\centering
\includegraphics[scale=1]{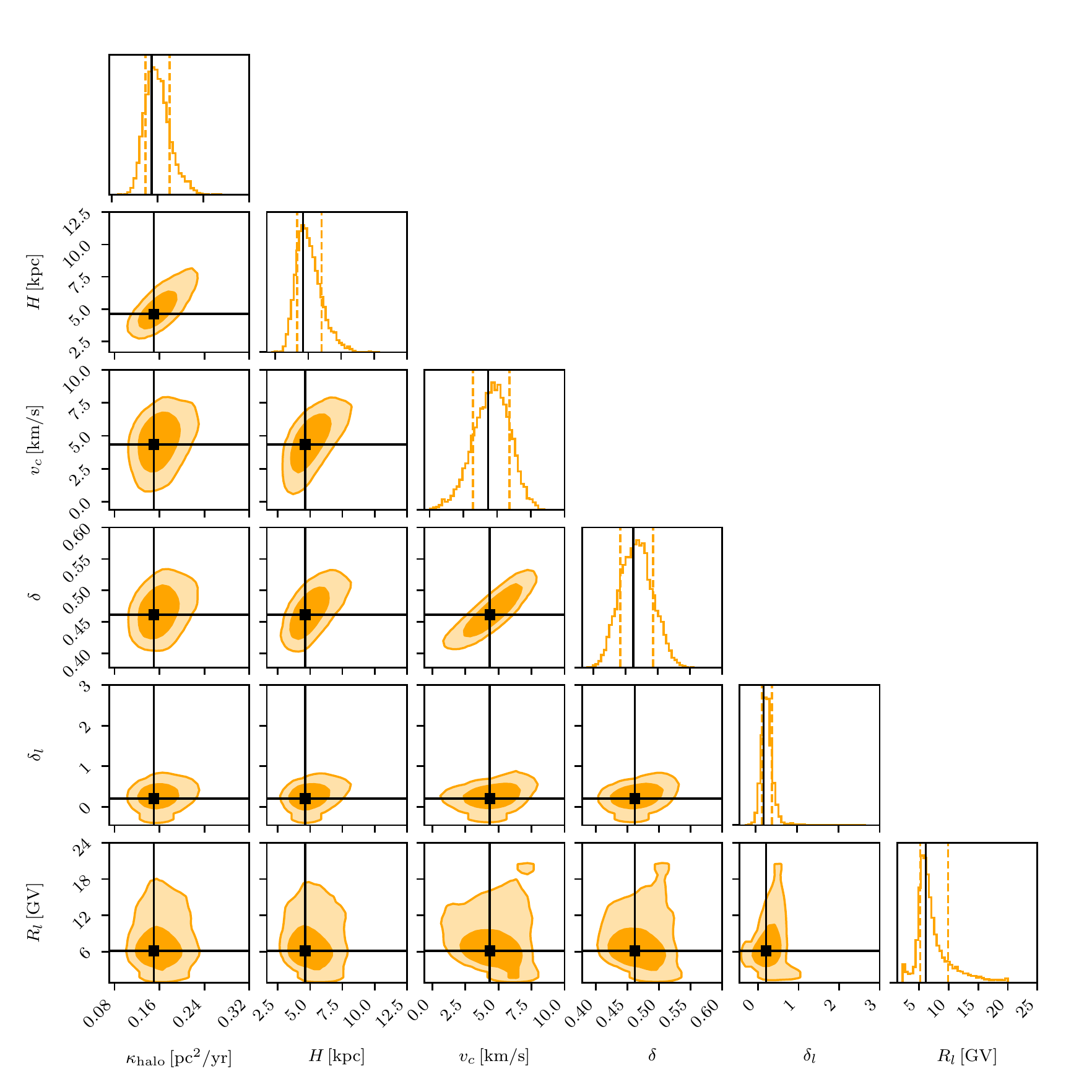}
\caption{
Same as Fig.~\ref{fig:corner_preliminary}, but for null model with existing data only.}
\label{fig:ncorner_nopre}
\end{figure*}

Removing preliminary AMS-02 data results in the corner plots shown in Fig.~\ref{fig:ncorner_nopre}. The degeneracy between $H$ and $\kappa_o$ still exist, but is less pronounced than in the full model. The halo height is similar to the value found by \cite{Wei2022}. Again the same volume effect as for the model with preliminary data is visible. 

\begin{figure*}
\centering
\includegraphics[scale=1]{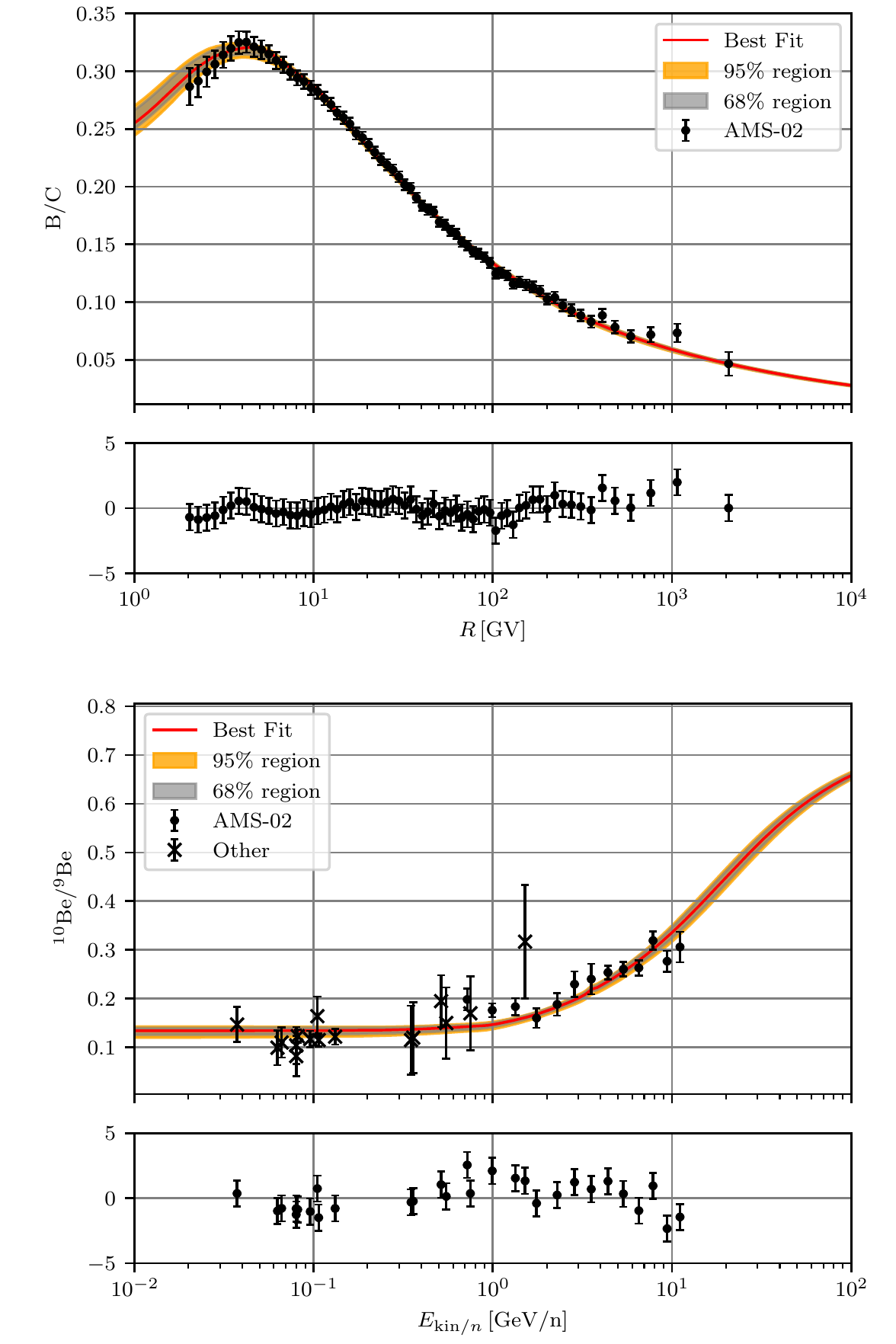}
\caption{Best fit spectra of B/C (top panel) and $\mathstrut^{10}\text{Be}/\mathstrut^{9}\text{Be}$ (bottom panel) including statistical $\unit{68}{\%} / \unit{95}{\%}$ quantiles for the full model ``w/ prelim.''. The red line indicates the best fit. The lower
panels are pull plots.}
\label{fig:spectra}
\end{figure*}